\documentclass[10pt,a4paper]{article}
\usepackage[utf8]{inputenc}
\usepackage{amsmath, amsthm, amssymb,mathrsfs}
\usepackage{enumerate}
\newtheorem{thm}{Theorem}[section]

\newtheorem{tvrzx}[thm]{Proposition}

\newtheorem{lemmax}[thm]{Lemma}

\newtheorem{theoremx}[thm]{Theorem}

\theoremstyle{definition}
\newtheorem{definicex}[thm]{Definition}

\theoremstyle{remark}
\newtheorem{remx}[thm]{Remark}

\theoremstyle{definition}
\newtheorem{examplex}[thm]{Example}
\newenvironment{example}{\begin{examplex}}{\medskip\end{examplex}}

\def\R{\mathbb{R}}

\def\<{\langle}
\def\>{\rangle}
\def\~{\widetilde}
\def\^{\wedge}

\def\io{\mathit{i}}

\def\D{\mathcal{D}}

\def\Te{\mathcal{T}}
\def\di{\mathsf{d}}
\def\cD{\nabla}

\def\cif{C^{\infty}(M)}

\newcommand{\TM}[1]{\Lambda^{#1} TM}
\newcommand{\cTM}[1]{\Lambda^{#1} T^{\ast}M}

\newcommand{\vf}[1]{ \mathfrak{X}^{#1}(M)}
\newcommand{\df}[1]{ \Omega^{#1}(M)}
\newcommand{\Li}[1]{ \mathcal{L}_{#1}}

\DeclareMathOperator{\Ric}{Ric}
\DeclareMathOperator{\Ann}{Ann}

\begin{document}
\begin{flushright}
\today
% preprint number (if any)
\end{flushright}
\vspace{0.7cm}
\begin{center}
 %\vskip1cm

\baselineskip=13pt {\Large \bf{Leibniz algebroids, generalized Bismut connections and Einstein-Hilbert actions}\\}
 \vskip1.3cm
 Branislav Jur\v co$^{1,2}$, Jan Vysoký$^{3,4}$\\
 \vskip0.6cm
$^{1}$\textit{Mathematical Institute, Faculty of Mathematics and Physics,
Charles University\\ Prague 186 75, Czech Republic, jurco@karlin.mff.cuni.cz}\\
\vskip0.3cm
$^{2}$\textit{Max Planck Institute for Mathematics, Vivatsgasse 7, 53111 Bonn, Germany}\\
\vskip0.3cm
$^{3}$\textit{Czech Technical University in Prague\\ Faculty of Nuclear Sciences
and Physical Engineering\\ Prague 115 19, Czech Republic, vysokjan@fjfi.cvut.cz}\\
\vskip0.3cm
$^{4}$\textit{Jacobs University Bremen\\ 28759 Bremen, Germany}\\
\vskip0.5cm
%email@email... \\
\end{center}

\begin{abstract}
Connection, torsion and curvature are introduced for general (local) Leibniz algebroids. 
Generalized Bismut connection on $TM\oplus \cTM{p}$ is an example leading to a scalar curvature of the form $R + H^2$ for a closed $(p+2)$-form $H$.
\end{abstract}

{\textit{Keywords:}} Generalized geometry, Leibniz algebroid, Bismut connection, torsion, curvature

\section{Introduction}
In this short note, we start to develop 
a general theory of connections, torsions and curvatures for local Leibniz algebroids. Interesting Leibniz algebroids are, for instance, those related to exceptional generalized geometries \cite{Hull}.  We believe that our constructions can be applied to a wide class of closed form Leibniz algebroids, classified in \cite{Baraglia}.
In all these Leibniz algebroids we have, in addition to the anchor controlling the Leibniz property in the second argument of the Dorfman bracket, also the so called locality operator controlling the behavior of the bracket under the multiplication of its first argument by a function. This locality operator can then be used to define the appropriate notions of torsion and curvature.

Though the general theory is simple and transparent, explicit computations are quite tedious even in the simplest examples. Hence, in this short note we present as an example only results for the simplest Leibniz algebroid on  $TM\oplus \cTM{p}$ equipped with the higher Dorfman bracket and with the corresponding generalized metric (defined by an ordinary Riemannian metric $g$ and by an $(p+1)$-form $C$). In this example, we generalize the (generalized) Bismut connection from the case $p=1$, whose significance in the context of generalized geometry was first
understood and investigated in \cite{Ellwood}. Its properties were highlighted in \cite{GualtieriBranes}, where its torsion was defined too. The calculations relating such metric connections with skew torsion to the Courant bracket go back to cite \cite{Hitchin,Gualtieri}. 

Of course, what we aim for are general definitions that in our example lead to the scalar curvature of the form  $R + (dC)^2$. 

There is a vast and important literature on supergravity actions from the point of view of (exceptional) generalized geometry and/or double field theory. It is far beyond the scope of this short note to comment on all of these, even to cite them. Among these, it seems to us that \cite{Waldram1, Waldram2} are, at least in some aspects, closest to our point of view and include an excellent overview of the literature.

\section{Local Leibniz algebroids}

Let us recall the notion of a Leibniz (Loday) algebroid.
A Leibniz algebroid is a triple $(E,\rho,\circ)$, where $E \stackrel{\pi}{\rightarrow} M$ is a (smooth) vector bundle, $\rho: E \rightarrow TM$ is a vector bundle morphism, called the anchor, and $\circ$ is an $\R$-bilinear bracket on sections $\Gamma(E)$ of $E$, satisfying the Leibniz rule
\begin{equation}
e \circ (fe') = f (e \circ e') + (\rho(e).f)e'
\end{equation}
and the Leibniz identity
\begin{equation}
e \circ (e' \circ e'') = (e \circ e') \circ e'' + e' \circ (e \circ e'')
\end{equation}
for all $e,e',e'' \in \Gamma(E)$ and $f\in C^{\infty}(M)$.

From the consistency of the Leibniz rule and the Leibniz identity under the replacement $e'' \mapsto fe''$, it follows that $\rho(e \circ e') = [\rho(e),\rho(e')]$. Further, we have a natural "differential" $\di$\footnote{not to be confused with ordinary de Rham differential on $M$} defined as a $\R$-linear map $\di: \cif \rightarrow \Gamma(E^{\ast})$ 
\begin{equation} \< \di f, e \> = \rho(e).f \end{equation}
for all $e \in \Gamma(E)$ and $f \in \cif$. Obviously $\di$ satisfies the usual Leibniz rule $\di(fg) = \di(f)g + f\di(g)$, for $f,g \in \cif$. 

Next, one can define a Lie derivative $\Li{}^{E}$, corresponding to $\circ$. It will define a first order differential operator on the tensor bundle $\Te(E)$. It is defined in the lowest orders and extended as a differential to all tensors.
On functions, it just the derivative in the direction of $\rho(e)$
\begin{equation} \Li{e}^{E}f = \rho(e).f \end{equation}
for all $e \in \Gamma(E)$ and $f \in \cif = \Gamma(\Te_{0}^{0}(E))$. 
On sections of $E$, it is the Leibniz bracket $\circ$ itself
\begin{equation} \Li{e}^{E}e' = e \circ e' \end{equation}
for all $e \in \Gamma(E)$ and $e' \in \Gamma(E) = \Gamma(\Te^{1}_{0}(E))$. 
On sections of $\Gamma(E^{\ast})$
\begin{equation} \label{eq_Lieon1forms} \< \Li{e}^{E}\alpha, e'\> = \Li{e}^{E}\<\alpha,e'\> - \<\alpha, \Li{e}^{E}e'\> \end{equation}
for all $f \in \cif$, $e,e' \in \Gamma(E)$ and $\alpha \in \Gamma(E^{\ast}) = \Gamma(\Te_{1}^{0}(E))$. 

Note that one has to use the Leibniz rule in order to guarantee the proper tensorial behavior on the right-hand side of the defining equations. For example, the right-hand side of (\ref{eq_Lieon1forms}) has to be $\cif$-linear in $e'$, which is guaranteed by Leibniz rule. 
On the other hand, Leibniz identity shows that $e \mapsto \Li{e}^{E}$ defines a bracket homomorphism
\begin{equation} \Li{e \circ e'}^{E} = \Li{e}^{E} \Li{e'}^{E} - \Li{e'}^{E} \Li{e}^{E} \end{equation}
for all $e,e' \in \Gamma(E)$. 

In general, one has no relation between $(fe) \circ e'$ and $e \circ e'$. As a consequence, $e \circ e'$ can depend on the values of the section $e$ at every point of the manifold $M$. If this happens, we can't restrict the bracket to local sections, which is necessary in order to write it in some local frame components. Hence, in the following we will restrict ourselves only to the so called local Lie algebroids, in particular the bracket $\circ$ will be a bidifferential operator of degree one.

We say that the Leibniz algebroid $(E,\rho,\circ)$ is a {\it local} one,\footnote{cf. Definition 4.7 and Theorem 4.8 in \cite{Grabowski}} if there exists $L \in \Gamma(\Te_{2}^{2}(E))$, such that
\begin{equation}
(fe) \circ e' = f (e \circ e') - (\rho(e').f)e + L(\di{f},e,e') 
\end{equation}
where $L$ is viewed as $\cif$-trilinear map $L: \Gamma(E^{\ast}) \times \Gamma(E) \times \Gamma(E) \rightarrow \Gamma(E)$. 

Obviously, $L$ in its first argument is defined uniquely only on the subbundle $\Ann(\ker{\rho})\subset E^*$, the annulator of the kernel of the anchor, which is locally generated by sections of the form $\di{f}$. Nevertheless, using the partition of unity we can define a scalar product on $E^\ast$ and extend $L$ trivially on the orthogonal complement to  $\Ann(\ker{\rho})$. Also, the $\cif$-trilinearity of $L$ is
essential for the definition to be a consistent one.

As a direct consequence of the definition of a  local Leibniz algebroid, we see that
\begin{equation} \label{eq_rhoonL}
\rho( L(\di{f},e,e') ) = 0 
\end{equation}
i.e., $L(\di{f},e,e')$ takes values in the subbundle $\ker{\rho}$. Moreover, we can always choose an $L$ satisfying $\rho( L(\beta,e,e') )=0$ for all $\beta \in \Gamma(E^*)$, $e,e'\in \Gamma(E)$, by extending it - as mentioned above - trivially to $((\Ann(\ker{\rho}))^\perp $.

\begin{example}
Lie algebroid is simply a Leibniz algebroid with skew-symmetric bracket. Obviously we can put $L = 0$ in this case.
\end{example}

\begin{example}\label{example1.2}
Courant algebroid is a Leibniz algebroid equipped with fiberwise metric $\<\cdot,\cdot\>_{E}$ satisfying
\begin{equation} \label{def_courant1} \<e', e \circ e \>_{E} = \frac{1}{2} \rho(e').\<e,e\>_{E} \end{equation}
for all $e,e' \in \Gamma(E)$ and being invariant with respect to the Lie derivative induced by $\circ$, that is
\begin{equation} \label{def_courant2} \rho(e).\<e',e''\>_{E} = \<e \circ e', e''\>_{E} + \<e' , e \circ e''\>_{E}
\end{equation}
for all $e,e',e'' \in \Gamma(E)$. Note that $\<\cdot,\cdot\>_{E}$ defines an isomorphism $\psi_{E}: E \rightarrow E^{\ast}$. Using this isomorphism, we can define yet another  differential $\D: \cif \rightarrow E$ as $\D = \psi_{E}^{-1} \circ \di$. The first axiom of a Courant algebroid can then be rewritten as
\begin{equation} e \circ e = \frac{1}{2} \D\<e,e\>_{E} \end{equation}
This can be polarized to 
\begin{equation} e \circ e' = - e' \circ e + \D\<e,e'\>_{E} \end{equation}
Hence,
\begin{equation} (fe) \circ e' = f (e \circ e') - (\rho(e').f)e + \<e,e'\>_{E} \D{f} \end{equation}
We thus get the condition on $L$ to be
\begin{equation} L(\di{f},e,e') = \<e,e'\>_{E} \psi_{E}^{-1}(\di{f}) \end{equation}
Therefore, in addition to the generic choice of $L$ satisfying $\rho( L(\beta,e,e') )=0$ we also have an another one $L(\beta,e,e') := \<e,e'\>_{E} \psi_{E}^{-1}(\beta)$, which seems to be more natural in case of a Courant algebroid.
\end{example}

\begin{example}\label{example1.3}
Consider $E = TM \oplus \cTM{p}$, with anchor $\rho = {\mbox pr}_{TM}$, the projection  to the tangent bundle $TM$, and bracket $\circ$ defined as
\begin{equation} \label{def_hdorfman} (x,a_{p}) \circ (y,b_{p}) = ([x,y], \Li{x}b_{p} - \io_{y}da_{p}) \end{equation}
for vector fields $x,y \in \vf{}$ and $p$-forms $a_{p},b_{p} \in \df{p}$. One can easily show that this is indeed a Leibniz algebroid. It is neither a Lie nor a Courant algebroid. One finds that
\begin{equation} (f(x,a_{p})) \circ (y,b_{p}) = f ((x,a_{p}) \circ (y,b_{p})) - (y.f)(x,a_{p}) + (0, df \^ \<(x,a_{p}),(y,b_{p})\>^{+}) \end{equation}
where $\<\cdot,\cdot\>^{+}$ is an $\df{p-1}$-valued pairing on $E$, defined as
\begin{equation} \<(x,a_{p}),(y,b_{p})\>^{+} = \io_{x}b_{p} + \io_{y}a_{p} \end{equation}
The dual bundle is $E^{\ast} = T^{\ast}M \oplus \TM{p}$, and the map $\di$ is then $\di{f} = (df, 0)$. Let $p_{1}: E^{\ast} \rightarrow T^{\ast}M$ be the projection onto the first factor. We find
\begin{equation} L(\di{f},e,e') = (0, p_{1}(\di{f}) \^ \<e,e'\>^{+}) \end{equation}
There is an obvious example of $L$ satisfying $\rho L(\beta,e,e')=0$. Namely,
\begin{equation} L(\beta,e,e') := (0, p_{1}(\beta) \^ \<e,e'\>^{+}) \end{equation}
for all $e,e' \in \Gamma(E)$ and $\beta \in \Gamma(E^{\ast})$. 
\end{example}

We have defined local Leibniz algebroids with the intention to be able write them in local coordinates/bases. Assume therefore that we have a neighbourhood $U$ with some local frame $(e_{1}, \dots, e_{k})$ for $E$, and a set of local coordinates $(y^{1}, \dots, y^{n})$ on $M$. We can define the structure functions of $\circ$ as
\begin{equation} e_{\alpha} \circ e_{\beta} = {c^{\lambda}}_{\alpha \beta} e_{\lambda} \end{equation}
The structure functions of $\rho$ are defined by
\begin{equation} \rho(e_{\alpha}) = \rho_{\alpha}^{k} \partial_{k} \end{equation}
and finally those of $L$ as 
\begin{equation} L(e^{\sigma},e_{\alpha},e_{\beta}) = {L^{\lambda \sigma}}_{\alpha \beta} e_{\lambda}
\end{equation}
If $e = v^{\alpha} e_{\alpha}$ and $e' = w^{\beta} e_{\beta}$, we can write the bracket of $e$ and $e'$ as 
\begin{equation} e \circ e' = \big( v^{\alpha} w^{\beta} {c^{\lambda}}_{\alpha \beta} + v^{\alpha} \rho_{\alpha}^{k} {w^{\lambda}}_{,k} - w^{\alpha} \rho_{\alpha}^{k} {v^{\lambda}}_{,k} + {v^{\alpha}}_{,k} w^{\beta}\rho^{k}_{\mu} {L^{\lambda \mu}}_{\alpha \beta} \big) e_{\lambda} \end{equation}
To simplify the notation, we introduce the partial derivative along $E$ directions, that is 
\begin{equation} f_{,\alpha}:=\partial_{\alpha}f :=  \rho_{\alpha}^{k} \partial_{k}f  =:\rho_{\alpha}^{k} f_{,k} = (\di{f})_{\alpha} \end{equation}
The above coordinate expression is then rewritten as
\begin{equation} e \circ e' = \big( v^{\alpha} w^{\beta} {c^{\lambda}}_{\alpha \beta} + v^{\alpha} {w^{\lambda}}_{,\alpha} - w^{\alpha} ( {v^{\lambda}}_{,\alpha} - {v^{\beta}}_{,\mu} {L^{\lambda \mu}}_{\beta \alpha}) \big) e_{\lambda} \end{equation}

\section{Linear connections on local Leibniz algebroids}
\subsection{Connection}
A local Leibniz algebroid $(E,\rho,\circ)$ is in particular a vector bundle with an anchor, a vector bundle morphism $\rho:E\to TM$. Hence, we can define a connection on it just mimicking the definition for the case of a Lie algebroid \cite{Fernandes}\footnote{For a thorough discussion of connections on Courant algebroids see \cite{Alekseev}}. 

Let $(E,\rho)$ be an anchored vector bundle and $V$ a vector bundle. We say that an $\R$-bilinear map $\cD: \Gamma(E) \times \Gamma(V) \rightarrow \Gamma(V)$ is a $E$-connection on $V$, if 
\begin{enumerate}
\item $\cD(fe,v) = f \cD(e,v)$
\item $\cD(e,fv) = f \cD(e,v) + (\rho(e).f)v$ 
\end{enumerate}
for all $e \in \Gamma(E), v \in \Gamma(V)$ and $f \in \cif$. We invoke the usual notation $\cD_{e} = \cD(e,\cdot)$, and call $\cD_{e}$ a covariant derivative along $e$.
If $E=V$ we say that $\nabla$ is a (linear) connection on $(E,\rho)$.
We say that an $E$-connection $\nabla$ on $V$ is induced by a $TM$-connection $\nabla'$ on $V$ if $\nabla_e = \nabla'_{\rho(e)}$.

Locally, in some frame $(e_{1}, \dots, e_{k})$, one can define Christoffel symbols by equation
\begin{equation}
\cD_{e_{\alpha}} e_{\beta} = {\Gamma^{\lambda}}_{\alpha \beta} e_{\lambda} 
\end{equation}
These are of course not tensors, and for $e = v^{\alpha}e_{\alpha}$ and $e' = w^{\beta}e_{\beta}$, we get
\begin{equation} \cD_{e}e' = v^{\alpha}( {w^{\lambda}}_{,\alpha} + {\Gamma^{\lambda}}_{\alpha \beta} w^{\beta} ) e_{\lambda} \end{equation}
We also can denote the covariant derivative using the usual semicolon formalism
\begin{equation} {w^{\lambda}}_{;\alpha} = {w^{\lambda}}_{\alpha} + {\Gamma^{\lambda}}_{\alpha \beta} w^{\beta} \end{equation}
Linear connections (covariant derivatives) can be extended to all tensors using standard formulas. In lowest orders
\begin{equation} \cD_{e}f := \rho(e).f, \ \<\cD_{e}\beta,e'\> := \cD_{e}\<e',\beta\> - \<\beta, \cD_{e}e'\> \end{equation}
for $e,e' \in \Gamma(E)$, $\beta \in \Gamma(E^{\ast})$ and $f \in \cif$. On higher tensors on $E$, $\cD$ is extended as usual. 

If we have at our disposal a fibre-wise metric $g_E$ on $E$, we say that the connection is metric compatible if
\begin{equation} \label{def_metriccomp}
\rho(e) g_E(e',e'') = g_E(\nabla_e e',e'')+ g_E( e',\nabla_e e'')
\end{equation}

\subsection{Torsion operator}
For a local Leibniz algebroid $(E,\rho,\circ,L)$, one would like to define a torsion. Obviously, the naive guess 
\begin{equation} T(e,e') = \cD_{e}e' - \cD_{e'}e - e \circ e' \end{equation}
does not work in general, it is not $\cif$-linear in $e$. Moreover, it is not antisymmetric in $(e,e')$. This is a minor drawback when compared to non-tensoriality. Once the non-tensoriality is fixed, the antisymmetrisation, if needed at all, is trivial. 

Here is our proposal. 
Let $\cD$ be a linear connection on a local Leibniz algebroid $(E,\rho,\circ,L)$ and let $e_{\lambda}$ be some (local) frame of $E$ and $e^{\lambda}$ the dual one. Then there is a well defined (not necessarily antisymmetric) torsion operator $T$ of the form
\begin{equation} \label{def_gentorsion} T(e,e') = \cD_{e}e' - \cD_{e'}e + L(e^{\lambda},\cD_{e_{\lambda}}e, e') - e \circ e' \end{equation}
A direct computation reveals that it is $\cif$-linear in $e$ and $e'$, and thus defines an element $T \in \Gamma(\Te_{2}^{1}(E))$.

Concerning the local expression for $T$, we define the components of $T$ as
\begin{equation} T(e_{\alpha},e_{\beta}) = {T^{\lambda}}_{\alpha \beta} e_{\lambda} \end{equation}
One finds that
\begin{equation}
{T^{\lambda}}_{\alpha \beta} = {\Gamma^{\lambda}}_{\alpha \beta} - {\Gamma^{\lambda}}_{\beta \alpha} + {\Gamma^{\sigma}}_{\mu \alpha} {L^{\lambda \mu}}_{\sigma \beta} - {c^{\lambda}}_{\alpha \beta}
\end{equation}

Note that this definition in particular includes the case of Courant algebroids. For Courant algebroids, there already exists definitions of torsion operator \cite{GualtieriBranes}, \cite{Alekseev}. The torsion introduced in \cite{GualtieriBranes} is the following one
\begin{equation} T'(e,e',e'') = \< \cD_{e}e' - \cD_{e'}e - \{e,e'\}, e''\>_{E} + \frac{1}{2}( \<\cD_{e''}e,e'\>_{E} - \< \cD_{e''}e', e\>_{E}) \end{equation}
where $ \{e,e'\}$ is the antisymmetrized Dorfman bracket (the Courant bracket).
A direct check shows the relation 
$T'(e,e',e'') = \frac{1}{2}( \<T(e,e'),e''\>_{E} - \<T(e',e),e''\>_{E})$, if one chooses $L$ given by
\begin{equation} \label{eq_torsionCourantLchoice} L(\beta,e,e') = \<e,e'\>_{E} \psi_{E}^{-1}(\beta) \end{equation}
Hence the torsion operator of \cite{GualtieriBranes} is a skew-symmetrized version of our torsion operator, with one index lowered by $\<\cdot,\cdot\>_{E}$. Note that the particular choice (\ref{eq_torsionCourantLchoice}) of $L$ was important in establishing the relation. In \cite{Alekseev}, a Courant algebroid torsion is defined to be a $3$-form $C$ given by
\begin{equation}
C(e,e',e'') = \frac{1}{3}\< \{e,e'\}, e'' \>_{E} - \frac{1}{2}\< \cD_{e}e' - \cD_{e'}e, e''\>_{E} + cyclic(e,e',e''). 
\end{equation}
For $\cD$ compatible with $\<\cdot,\cdot\>_{E}$ in the sense of (\ref{def_metriccomp}), one can prove that in fact $C = -T'$. 

Another notion of the generalized torsion, which can be applied to Leibniz algebroids, was introduced in \cite{Waldram1, Waldram2}. They defined it as difference of "covariantized" Dorfman derivative and ordinary Dorfman derivative. By Dorfman derivative they mean the Lie derivative $\Li{}^{E}$ with respect to the first section. By "covariantized" derivative they mean Lie derivative with partial derivatives in coordinate expression replaced by covariant derivatives. We can relate this to our torsion in a holonomic frame,
\[ e_{\alpha} \circ e_{\beta} = 0 \]
in which case
\begin{equation} (\Li{e}^{E}e')^{\lambda} = v^{\alpha} {w^{\lambda}}_{,\alpha} - w^{\alpha} ( {v^{\lambda}}_{,\alpha} - {v^{\beta}}_{,\mu} {L^{\lambda \mu}}_{\beta \alpha}) 
\end{equation}
for $e=v^\alpha e_\alpha$ and $e'=w^\alpha e_\alpha$.
The covariantized Lie derivative is thus 
\begin{equation} (\Li{e}^{\cD}e')^{\lambda} = v^{\alpha} {w^{\lambda}}_{;\alpha} - w^{\alpha} ( {v^{\lambda}}_{;\alpha} - {v^{\beta}}_{;\mu} {L^{\lambda \mu}}_{\beta \alpha}) 
\end{equation}
Hence, according to the definition of \cite{Waldram1, Waldram2}
\begin{equation} \label{eq_torsionasdifference}
T(e,e') = (\Li{e}^{\cD} - \Li{e}^{E})e' \end{equation}
Note that $\Li{e}^{\cD}e'$ can be rewritten as
\begin{equation} \Li{e}^{\cD}e' = \cD_{e}e' - \cD_{e'}e + L(e^{\mu}, \cD_{e_{\mu}}e,e') \end{equation}
Now it is obvious that (\ref{eq_torsionasdifference}) gives exactly the formula (\ref{def_gentorsion}). Had we not assumed the holonomicity of the frame, $T(e,e')$ in (\ref{eq_torsionasdifference}) would contain, compared with our definition, an additional $v^{\alpha} w^{\beta}{c^{\lambda}}_{\alpha \beta}$ term, and would not be a well-defined section of $\Gamma(E)$ anymore. 

Obviously, for induced connections, their torsion  (\ref{def_gentorsion}) doesn't depend on choice of $L$.

\subsection{Curvature operator}
Here, we would like to define a curvature operator for connections on local Leibniz algebroids. As for the torsion operator, the first naive guess would be
\begin{equation}
R(e,e')e'' = \cD_{e}\cD_{e'}e'' - \cD_{e'}\cD_{e}e'' - \cD_{e \circ e'}e''
\end{equation}
for all $e,e',e'' \in \Gamma(E)$. Due to the property of the anchor $\rho(e\circ e')=[\rho(e),\rho(e')]$, such an $R(e,e')$ is $\cif$-linear in $e''$, i.e. a vector bundle morphism. As easily identified, the problem lies in the $\cif$-linearity in $e$ (the $\cif$-linearity in $e'$ is more or less obvious from the definitions). Here is our proposal how to fix this: 

Let $(E,\rho,\circ,L)$ be a local Leibniz algebroid, such that $\rho \circ L = 0$. Let $\cD: E \rightarrow \D(E)$ be a linear connection on $E$. The formula 
\begin{equation} \label{eq_Rdef} R(e,e')e'' = \cD_{e}\cD_{e'}e'' - \cD_{e'}\cD_{e}e'' + \cD_{L(e^{\alpha},\cD_{e_{\alpha}}e,e')}e'' - \cD_{e \circ e'}e'' 
\end{equation}
is $\cif$-linear in $e,e',e''$ and thus defines an element $R \in \Gamma(\Te_{3}^{1}(E))$. We call $R$ the curvature operator, or when viewed as a tensor, we call it $R$ the Riemann tensor of the linear connection $\cD$. Note that the condition $\rho \circ L = 0$ is essential in order not to destroy the tensoriality in $e''$. It is a straightforward check to see that the additional term $\cD_{L(e^{\alpha},\cD_{e_{\alpha}}e,e')}e''$ indeed cancels the nontensoriality in $e$. It preserves the tensoriality in $e'$ though.

In coordinates, one defines components of $R$ as
\begin{equation} R(e_{\alpha},e_{\beta})e_{\mu} = ({R^{\lambda}}_{\mu \alpha \beta} ) e_{\lambda}
\end{equation}
and one finds the explicit expression for those
\begin{equation}
\begin{split}
{R^{\lambda}}_{\mu \alpha \beta} & = {\Gamma^{\lambda}}_{\beta \mu,\alpha} - {\Gamma^{\lambda}}_{\alpha \mu,\beta} + {\Gamma^{\lambda}}_{\alpha \kappa} {\Gamma^{\kappa}}_{\beta \mu} - {\Gamma^{\lambda}}_{\beta \kappa} {\Gamma^{\kappa}}_{\alpha \mu} + {\Gamma^{\delta}}_{\sigma \alpha} {L^{\kappa \sigma}}_{\delta \beta} {\Gamma^{\lambda}}_{\kappa \mu} - {c^{\kappa}}_{\alpha \beta} {\Gamma^{\lambda}}_{\kappa \mu}
\end{split}
\end{equation}

Note that $R(e,e')$ is not necessarily skew-symmetric in $(e,e')$. Of course, we can always skew-symmetrize it in $(e,e')$
\begin{equation}R_{a}(e,e')e'' = \{ \cD_{e}\cD_{e'} - \cD_{e'}\cD_{e} + \frac{1}{2}\cD_{L(e^{\alpha},\cD_{e_{\alpha}}e,e')} - \frac{1}{2} \cD_{L(e^{\alpha},\cD_{e_{\alpha}}e',e)} - \cD_{\{e,e'\}} \}e'' \end{equation}

Ricci tensor is defined as usual 
$$\Ric (e,e') = \langle e^\alpha, R(e,e_\alpha)e'\rangle$$
 
For a metric compatible connection we define the Ricci scalar in a standard way as
$$\mathcal{R} = \Ric (g^{-1}e^\alpha,e_\alpha)$$

Let us note that for an induced connection, the term containing the operator $L$ doesn't contribute to $R$ at all and we have a more traditionally looking expression of the form 
$$R(e,e')e''= (\nabla_{\rho(e)} \nabla_{\rho(e')}-\nabla_{\rho(e')} \nabla_{\rho(e)} - \nabla_{[\rho(e),\rho(e')]})e''$$

\section{Generalized Bismut connection}
\subsection{Connection}
In \cite{Ellwood} a generalized Bismut connection on the Courant algebroid of Example \ref{example1.2} was introduced. Here, we generalize it to the case of Example \ref{example1.3}. Hence, we consider the local Leibniz algebroid $E = TM \oplus \cTM{p}$ with its higher Dorfman bracket and the map $L$ chosen as\footnote{Obviously, for the connection, the existence of the Dorfman bracket and locality are not necessary.} 
\begin{equation} \label{LforBismut}
L(\beta,e,e') = (0, p_{1}(\beta) \^ \<e,e'\>^{+}) 
\end{equation}
Such an $L$ satisfies $\rho \circ L = 0$. 
We define a generalized metric $\mathbf{G}$ and a connection $\nabla$ compatible with this metric. Let $g$ be a metric on $M$ and  $\~g$ is a skew-symmetrized $p$-fold tensor product of $g$, defining a fiberwise metric on $\TM{p}$. Also, let $C \in \df{p+1}$ be a $(p+1)$-form on $M$. With an abuse of notation, we introduce the corresponding maps $g:\vf{} \rightarrow \df{1} $, ${\~g}^{-1}: \df{p}\rightarrow \vf{p}$, $C: \vf{p} \rightarrow \df{1}$. For instance,
\[ C(q) = C_{iJ} q^{J} dy^{j} \]
with the transpose map $C^{T}: \vf{} \rightarrow \df{p}$ being an insertion of vector field into $(p+1)$-form $C$: $C^{T}(x) = \io_{x}C$ for all $x \in \vf{}$. Then $\mathbf{G}$ is defined as
\begin{equation} \mathbf{G} = \begin{pmatrix} 1 & C \\ 0 & 1 \end{pmatrix}
\begin{pmatrix} g & 0 \\ 0 & \~g^{-1} \end{pmatrix}
\begin{pmatrix} 1 & 0 \\ C^{T} & 1 \end{pmatrix}
\end{equation}
It maps $\vf{}\oplus \df{p}$ to  $\df{}\oplus \vf{p}$.

If $\cD_{x}^{LC}$ is the Levi-Civita connection on $M$ and $H:= dC$, then the generalized Bismut connection $\cD$ will be defined so that the covariant derivative 
$\cD_{(x,a_p)}$ will not depend on the $p$-form $a_p$, which will be indicated as
$\cD_{(x,0)}$. Hence, the generalized Bismut connection will be an induced one. Before giving the definition, we introduce another, a bit simpler, connection $\widehat\cD$ related to $\cD$ as
\begin{equation}
\cD_{(x,0)} = e^{C} \widehat\cD_{(x,0)} e^{-C},
\end{equation}
where the map $e^{C}$ is defined as
\begin{equation}
e^{C}(y,b_{p}) := (y, b_{p} - \io_{y}C) = \begin{pmatrix} 1 & 0 \\ -C^{T} & 1 \end{pmatrix} \begin{pmatrix} y \\ b_{p} \end{pmatrix}
\end{equation}
The expression for the connection $\widehat\cD$ is 
\begin{equation}
\widehat\cD_{(x,0)}=
\begin{pmatrix}
\cD_{x}^{LC} & - \frac{1}{2}g^{-1}H(x, \~g^{-1}(\star),\cdot) \\
-\frac{1}{2}H(x,\star,\cdot) & \cD_{x}^{LC} 
\end{pmatrix}
\end{equation}
where $\star$ indicates places where components of the pair $(y,b_p)\in \vf{}\oplus \df{p}$ acted upon by the covariant derivative are inserted.
Explicitly, when acting on $(y,0)$ we have 
\[\widehat\cD_{(x,0)}(y,0)=(\cD_{x}^{LC}y,  -\frac{1}{2}H(x,y,\cdot))\]
and when acting on $(0,b_p)$ we obtain
\[\widehat\cD_{(x,0)}(0,b_p)=(- \frac{1}{2}g^{-1}H(x, \~g^{-1}(b_p),\cdot),\cD_{x}^{LC} b_p)\]

It is a rather straightforward check that the connection $\widehat\cD$ is compatible with the generalized metric $\widehat{\mathbf{G}} := {\mbox {diag}}(g, \tilde{g}^{-1})$. 
Hence, as a consequence, the generalized Bismut connection $\cD$ is compatible with the generalized metric $\mathbf{G}$. 
\subsection{Torsion operator} \label{subsec_torsionhbismut}
Here, we write down the result of a rather lengthy and tedious computation of the torsion operator of the generalized Bismut connection $\cD$. We split the result into the vector field and $p$-form  components $T_{1}$ and $T_{2}$, respectively.
However, let us start with formulas for the respective torsion components $\widehat T_{1}$ and $\widehat T_{2}$ of the simpler connection  $\widehat\cD$. We have 
\begin{equation}
\widehat T_{1}((x,a_{p}),(y,b_{p})) = \frac{1}{2}g^{-1} \big( H(y,\~g^{-1}(a_{p}),\cdot) - H(x,\~g^{-1}(b_{p}),\cdot) \big)
\end{equation}
for the fist one and
\begin{equation}
\widehat T_{2}((x,a_{p}),(y,b_{p})) = - \frac{3}{2}H(x,y,\cdot) - \frac{1}{2}e^{k} \^ \io_{g^{-1}H(e_{k},\~g^{-1}(a_{p}),\cdot)}b_{p}
\end{equation}
for the second one. Here $e_k$ and $e^k$ are elements of some mutually dual frames of $TM$ and $T^\ast M$, respectively.
To get a better understanding how the second form in this formula works, contract it against a $p$-tuple of vector fields $(z_{1},\dots,z_{p})$. The result will be
\begin{equation} - \frac{1}{2}( e^{k} \^ \io_{g^{-1}H(e_{k},\~g^{-1}(a_{p}),\cdot)}b_{p})(z_{1},\dots,z_{p}) = - \sum_{i=1}^{p} \frac{1}{2} b_{p}(z_{1}, \dots, g^{-1}H(z_{i},\~g^{-1}(a_{p}),\cdot), \dots, z_{p}). \end{equation}
Note, for $p > 1$, the torsion is not necessarily skew-symmetric in $(e,e')$. 

The relation between the the hatted and unhatted torsions is\footnote{For the relation it is important that $\widehat\nabla_e = \widehat\nabla_{e^{-C}e}$, since  $\widehat\nabla_e$ depends only on the vector field part of $e$, and the well known property of the Dorfman bracket $e^{-C}(e^C e \circ e^C e') = e\circ e' + (0,H(\rho(e),\rho(e'),.))$.}
\begin{equation}
T(e,e') = e^{C}( \widehat T(e^{-C}e,e^{-C}e') + (0, H(\rho(e),H(\rho(e'))). 
\end{equation}
and we get 
\begin{equation}
T_{1}((x,a_{p}),(y,b_{p}))  
= \frac{1}{2}g^{-1}(H(y,\~g^{-1}(a_{p} + \io_{x}C),\cdot) - H(x,\~g^{-1}(b_{p}+\io_{y}C), \cdot)) 
\end{equation}
and 
\begin{equation} 
\begin{split}
T_{2}((x,a_{p}),(y,b_{p})) & =  -\frac{1}{2}H(x,y,\cdot) - \frac{1}{2} e^{k} \^ \io_{g^{-1}H(e_{k}, \~g^{-1}(a_{p} + \io_{x}C),\cdot)} (b_{p} + \io_{x}C) \\
& - \frac{1}{2} \io_{g^{-1}(H(y,\~g^{-1}(a_{p} + \io_{x}C),\cdot) - H(x,\~g^{-1}(b_{p} + \io_{y}C),\cdot))}C 
\end{split}
\end{equation}
We finish this subsection with the comment on $p=1$ case. The generalized Bismut connection in this case is same as the one in \cite{Ellwood}. Also, our torsion is for $p=1$ the same as the one of \cite{GualtieriBranes}. Since the connection is an induced one, both natural choices for $L$ have to give the same torsion.
Our choice (\ref{LforBismut}) of $L$ not only works for $p>1$, as we have seen, its property $\rho \circ L = 0$, also is essential for our definition of curvature to work.

\subsection{Curvature operator}
Here, we give the result of calculation of the Ricci scalars of the connections $\widehat\cD$ and $\cD$, they will turn out to be the same.  We start with the simpler primed connection $\widehat\cD$. Again, we split the result into the vector field and the $p$-form parts $\widehat R_1$ and $\widehat R_2$, respectively. 
\begin{align}
\widehat R_1((x,a_{p}),(y,b_{p}))(z,c_{p}) & = R^{LC}(x,y)z - \frac{1}{2}g^{-1}((\cD_{x}^{LC}H)(y,\~g^{-1}(c_{p}),\cdot) - (\cD_{y}^{LC}H)(x,\~g^{-1}(c_{p}),\cdot)) \nonumber \\
& + \frac{1}{4} g^{-1}(H(x,\~g^{-1}H(y,z,\cdot),\cdot) - H(y,\~g^{-1}H(x,z,\cdot),\cdot))
\end{align}
and
\begin{equation}
\begin{split}
\widehat R_2((x,a_{p}),(y,b_{p})(z,c_{p}) & = R^{LC}(x,y)c_{p} - \frac{1}{2} (\cD_{x}^{LC}H)(y,z,\cdot) + \frac{1}{2} (\cD_{y}^{LC}H)(x,z,\cdot).
\end{split}
\end{equation}
For the Ricci tensor, the only component contributing nontrivially to the Ricci scalar is
\begin{equation}
{\widehat\Ric}((x,0),(z,0)) = \Ric^{LC}(x,z) + \frac{1}{4}H(x, \~g^{-1}H(e_{k},z,\cdot), g^{-1}(e^{k}))
\end{equation}

The scalar curvature is defined using the fiberwise metric $\widehat{\mathbf{G}}$. We get
\begin{equation}
\begin{split}
\widehat{\mathcal{R}} & = \mathcal{R}^{LC} + \frac{1}{4}H(g^{-1}(e^{l}), \~g^{-1}H(e_{k},e_{l},\cdot), g^{-1}(e^{k})) = \mathcal{R}^{LC} + \frac{1}{4} g^{im} g^{kn} \~g^{IJ} H_{mIn} H_{kiJ} \\
& = \mathcal{R}^{LC} + \frac{(-1)^{p+1}}{4} H_{nmJ} H^{nmJ} 
\end{split}
\end{equation}

Again due to the fact that $\widehat{\nabla}_e$ depends only on the vector field part of $e$ it is easy to find the following relations between primed and unprimed curvatures
\begin{equation}
R(e,e')e'' = e^C\{\widehat R(e^{-C}e,e^{-C}e')e^{-C}e''\}  
\end{equation}
Now we can compute the Ricci scalar $\mathcal R$ using the generalized metric $\mathbf G$. From the above relation in follows that 
\begin{equation}
{\mathcal R}=\widehat{\mathcal R}=\mathcal{R}^{LC} + \frac{(-1)^{p+1}}{4} H_{nmJ} H^{nmJ} 
\end{equation}
leading to the generalized Einstein-Hilbert action
$$ S =\int \sqrt{g}(\mathcal{R}^{LC} + \frac{(-1)^{p+1}}{4} H_{nmJ} H^{nmJ}) $$

\section*{Acknowledgement}
It is a pleasure to thank Peter Schupp for discussions.
The research of B.J. was supported by grant GA\v CR P201/12/G028.
The research of J.V. was supported by Grant Agency of the Czech
Technical University in Prague, grant No. SGS13/217/OHK4/3T/14. J.V. also
gratefully acknowledges support from the DFG within the Research Training Group 1620 ``Models of Gravity''.

\end{document}